\documentclass{article}
\usepackage{graphicx}
\usepackage{amsmath}

\begin{document}
\title{Singularities on charged viscous droplets}
\author{{ S. I. Betel\'{u}}\\
Department of Mathematics\\
University of North Texas\\
P.O. Box 311430, Denton, TX 76203-1430\\\\
{ M. A. Fontelos}\\
Facultad de Ciencias\\ 
Universidad Aut\'{o}noma de Madrid\\ 
28049 Madrid, Spain\\\\
{U. Kindel\'{a}n}\\
Departamento de Matem\'atica Aplicada,\\
Universidad Polit\'ecnica de Madrid,\\
Rios Rosas 21,\\ 28003 Madrid Spain\\\\
O. Vantzos\\
Department of Mathematics\\
University of North Texas\\
}

\maketitle

\begin{abstract}
We study the evolution of charged droplets of a conducting viscous liquid.
The flow is driven by electrostatic repulsion and capillarity. These
droplets are known to be linearly unstable when the electric charge is above
the Rayleigh critical value. Here we investigate the nonlinear evolution
that develops after the linear regime. Using a boundary elements method, we
find that a perturbed sphere with critical charge evolves into a fusiform
shape with conical tips at time $t_0$, and that the velocity at the tips
blows up as $(t_0-t)^\alpha$, with $\alpha$ close to $-1/2$.
In the neighborhood of the singularity, the shape of the surface is
self-similar, and the asymptotic angle of the tips is smaller than the
opening angle in Taylor cones.
\end{abstract}

One of the leading problems in fluid dynamics is the formation of
singularities on charged masses of fluid.
These problems are relevant in a variety of physical and technological
situations, such as the breakup of water droplets in thunderstorms,
electrospraying and electropainting.

Here we provide evidence of the formation of finite time singularities
on electrically charged droplets of a
conducting viscous fluid immersed in a dielectric viscous fluid of
infinite extent. These singularities appear as the finite-time formation of
conical tips at the surface and the blow-up of the velocity field at that
point.

The interest in the shape of electrified drops dates back to Lord Rayleigh
\cite{Rayleigh}, who showed that if the electric charge is larger
than some critical value, then a spherical drop becomes unstable. For a drop
with total charge $Q$, surface tension coefficient $\gamma $ and radius $R$
suspended in a medium of dielectric constant $\varepsilon _{0} $, this
critical value is $Q_c=\sqrt{32\gamma\pi^2\varepsilon_0 R^3}$.
After the drop becomes unstable, it desintegrates into droplets of smaller
size. However, in recent experiments (see \cite{Duft}) it has been noticed
that previous to drop disintegration, the drop evolves into a
prolate spheroid which, after a finite time, develops conical tips from
which thin jets emerge.

This emission has been previously observed from a meniscus by Taylor, and is
the basis of cone-jet electrospraying, a technique to produce microdroplets
in a controlled fashion. A related technique, electrospray ionization, has
revolutionized mass spectroscopy of large molecules. Applications to
micro/nano encapsulation have also been developed recently
(cf. \cite{barrero}), and also the development of micro thrusters for
the propulsion of spacecraft using liquid jets \cite{GCH}.

Here, by means of a numerical calculation and asymptotic analysis, we find
that droplets with Rayleigh's critical charge evolve into fusiform shapes
with cones at the tips. On the course of this evolution, both the curvature
and the fluid velocity at the tips diverge as $(t_{0}-t)^\alpha$, where
$t_{0}$ is the time at which the cones are formed and the exponent $\alpha$
is approximately $-1/2$. Moreover, the numerical
experiments indicate that the local shape of the surface is self-similar.
The semiangle of the conical tips depends on
the contrast of viscosities between the outer and inner fluids, and is
close to $25$ degrees.

The existence of cones at the surface of a charged fluid is known since
the work of Taylor (cf. \cite{Taylor}), who computed {\it static}
solutions at the surface of a charged conducting fluid under the
influence of an electric field. These solutions are nowadays known as Taylor cones, and
possess a typical opening semiangle of $49.3^{o}$. The cones that we study
here are different from the Taylor cones since they are dynamic. We shall
refer to them as \textit{dynamic Taylor cones}.
A conclusion of our study is that static Taylor cones are not the generic
singular structures developing during the evolution of the surface of a
charged drop.
In future publications we will analyze how the dynamic
solutions can be continued after the formation of cones and determine
whether jets or a string of droplets are produced.

We assume that the drop occupies a region $\Omega (t)$. Since the drop is a
conductor, all the electric charge will be located at the boundary $\partial
\Omega $, and since the surrounding medium is a dielectric, the total charge
$Q$ remains constant. The electric field $\mathbf{E}$ outside the drop is
given by $\mathbf{E}=-\nabla V$ where $\Delta V=0\;\;\mbox{ in }%
R^{3}\backslash \Omega ,$\textbf{\ }$V=C\;\;\mbox{ on }\partial \Omega $%
\textbf{\ }and $V$\ decays at infinity. At the surface of a conductor, the
surface charge density $\sigma $ is given by the normal derivative of the
potential, $\sigma =-\varepsilon _{0}\frac{\partial V}{\partial n}$,\ so
that the repulsive electrostatic force per unit area is
\begin{equation}
\mathbf{F_{e}}=\frac{\mathbf{E}\sigma }{2}=\frac{\varepsilon _{0}}{2}\left(
\frac{\partial V}{\partial n}\right) ^{2}\mathbf{n}=\frac{\sigma ^{2}}{%
2\varepsilon _{0}}\mathbf{n}  \label{rep}
\end{equation}
where $\mathbf{n}$\ is the outward normal to the surface.

The fluid velocity $\mathbf{u}$ and the fluid pressure $p$ inside the drop
satisfy the Stokes equations
\begin{eqnarray}
-\nabla p+\mu_1\Delta \mathbf{u}=0 \;\;\mbox{in } \Omega(t),  \label{ff51} \\
\nabla \cdot \mathbf{u}=0 \;\;\mbox{in }\Omega (t)  \label{ff52}
\end{eqnarray}
where $\mu_1$ is the viscosity of the liquid inside the drop. Equations
similar to (\ref{ff51}) and (\ref{ff52}) must be satisfied by the
velocity and the pressure outside of the drop, $\mathbf{R}^{3}\backslash
\Omega(t)$, with $\mu_1$ replaced by $\mu_2$, the viscosity of the
surrounding liquid.

The boundary condition for the stress is
\begin{equation}
(T^{(2)}-T^{(1)})\mathbf{n}=\left( \gamma \kappa -\frac{\sigma ^{2}}{%
2\varepsilon _{0}}\right) \mathbf{n}\;\;\mbox{ on }\partial \Omega (t),
\label{ff53}
\end{equation}%
where $\kappa $ is the mean curvature of the surface and $T^{(k)}$ is the
stress tensor inside ($k=1$) or outside ($k=2$) the drop, given by
\begin{equation}
T_{ij}^{(k)}=-p\delta _{ij}+\mu _{k}\left( \frac{\partial u_{i}}{\partial
x_{j}}+\frac{\partial u_{j}}{\partial x_{i}}\right) \;\mbox{, }k=1,2\;.
\label{ff55}
\end{equation}
Equation (\ref{ff53}) expresses the balance between viscous stress,
capillary forces and electrostatic repulsion.
The kinematic condition is
\begin{equation}
v_{N}=\mathbf{u}\cdot \mathbf{n}\;\;\mbox{ on }\partial \Omega (t)
\label{ff54}
\end{equation}%
where $v_{N}$ is the normal velocity of the free boundary $\partial \Omega
(t)$.

Our numerical method to compute the evolution of the drop is based on the
boundary integral method for the Stokes system (see \cite{Pozrikidis},
\cite{Acrivos} for a comprehensive explanation). In this method, the equation for
the velocity at $\partial \Omega (t)$ is written in integral form as
\begin{eqnarray}
u_{j}(\mathbf{r}_{0}) &=&-\frac{1}{4\pi }\frac{1}{\mu _{1}+\mu _{2}}%
\int_{\partial \Omega (t)}f_{i}(\mathbf{r})G_{ij}(\mathbf{r},\mathbf{r}%
_{0})dS(\mathbf{r})   \\
&&-\frac{1}{4\pi }\frac{\mu _{2}-\mu _{1}}{\mu _{2}+\mu _{1}}\int_{\partial
\Omega (t)}u_{i}(\mathbf{r})T_{ijk}(\mathbf{r},\mathbf{r}_{0})n_{k}(\mathbf{r%
})dS(\mathbf{r})\;. \notag \label{uk}
\end{eqnarray}
where
\begin{eqnarray}
G_{ij}(\mathbf{r},\mathbf{r}_{0}) &=&\frac{\delta _{ij}}{\left\vert \mathbf{r%
}-\mathbf{r}_{0}\right\vert }+\frac{(r_{i}-r_{0,i})(r_{j}-r_{0,j})}{%
\left\vert \mathbf{r}-\mathbf{r}_{0}\right\vert ^{3}} \\
T_{ijk}(\mathbf{r},\mathbf{r}_{0}) &=&-6\frac{%
(r_{i}-r_{0,i})(r_{j}-r_{0,j})(r_{k}-r_{0,k})}{\left\vert \mathbf{r}-\mathbf{%
r}_{0}\right\vert ^{5}} \\
f_{i}(\mathbf{r}) &=&\left[ \gamma (\mathbf{r})-\frac{\varepsilon _{0}}{2}%
\left( \frac{\partial V}{\partial n}\right) ^{2}(\mathbf{r})\right] n_{i}
(\mathbf{r})  .
\end{eqnarray}
The equation for the charge density is
\begin{eqnarray}
V(\mathbf{r}_{0}) = \frac{1}{4\pi } \int_{\partial \Omega (t)} \frac{\sigma(%
\mathbf{r})}{|\mathbf{r-r_0}|} dS(\mathbf{r}).  \label{intv}
\end{eqnarray}
This integral equation must be inverted numerically to obtain the charge
density. $V(\mathbf{r}_{0})$ is a constant along the surface, and it is
determined by the condition
\begin{equation}
Q = \int_{\partial \Omega (t)} \sigma(\mathbf{r})dS(\mathbf{r}).
\label{cargav}
\end{equation}

Then we discretize the axisymmetric surface with $N$ conical rings and the
velocity and the potential are approximated by constants within each ring.
This leads to a system of linear equations that is solved using the LU decomposition.
Our method increases its stability by applying a singularity removal
procedure as explained in section 6.4 (formula 6.4.3) of \cite{Pozrikidis}.
As we are restricted to axisymmetric configurations, the surface integrals
in (\ref{uk}) transform then into line integrals with kernels given in terms of elliptic
functions (see section 2.4 in \cite{Pozrikidis}). We validated the numerical
scheme by comparing the numerical solutions with the linearized theory for
viscous drops, that describes the evolution of small amplitude deformations.

As the initial condition, we start with a spherical droplet perturbed with a
spherical harmonic $r(\theta)= R+\epsilon Y_2^0(\theta)$,
where $r, \theta$ are the spherical coordinates with origin at the center of
the drop. We found that the results are not sensitive on the value of
$\epsilon$, as long as it is smaller than $0.1$. Here we will show the
results for this value of $\epsilon$.

In figure 1 we show the drop profiles at several times approaching the
singularity, that shows as conical tips at both ends of the drop.
\begin{figure}[tbp]
\centering
\includegraphics[width=0.75\textwidth]{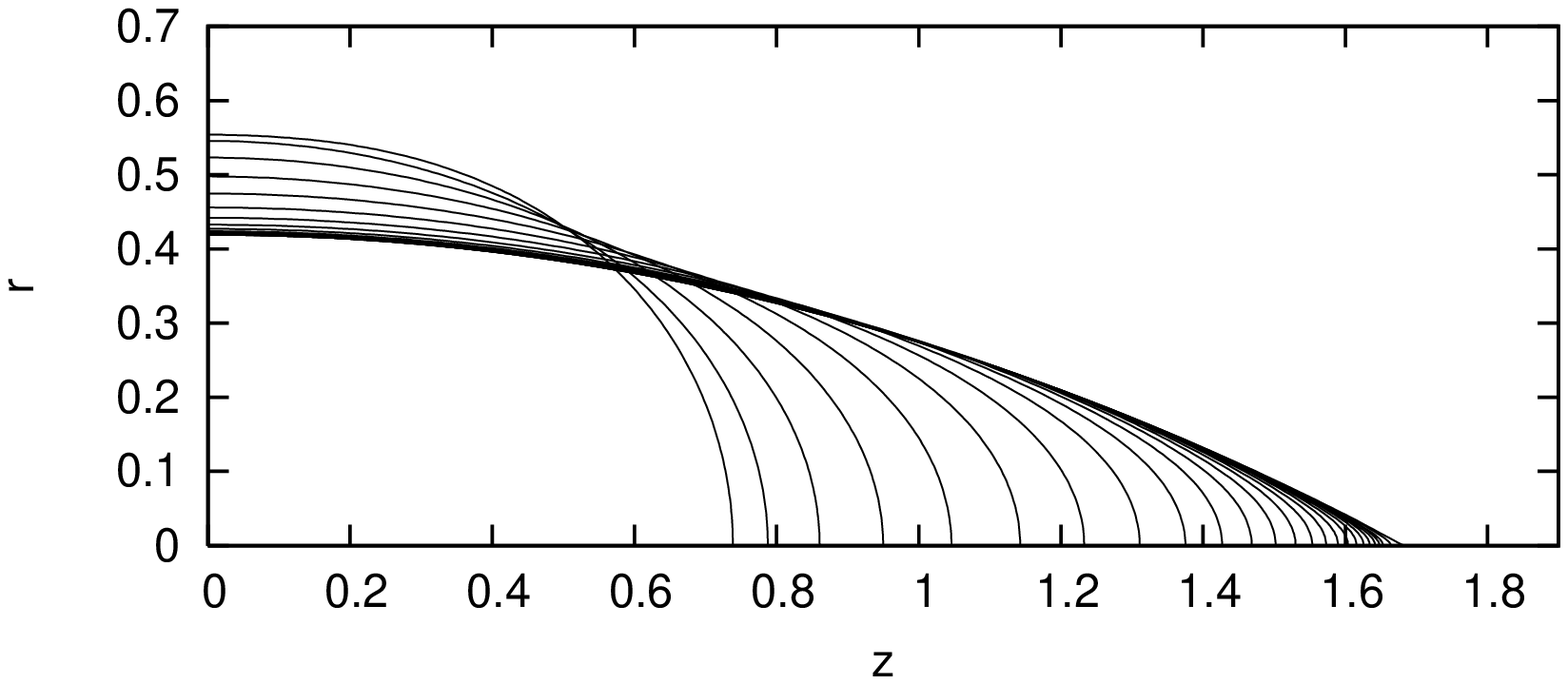} 
\includegraphics[width=0.75\textwidth]{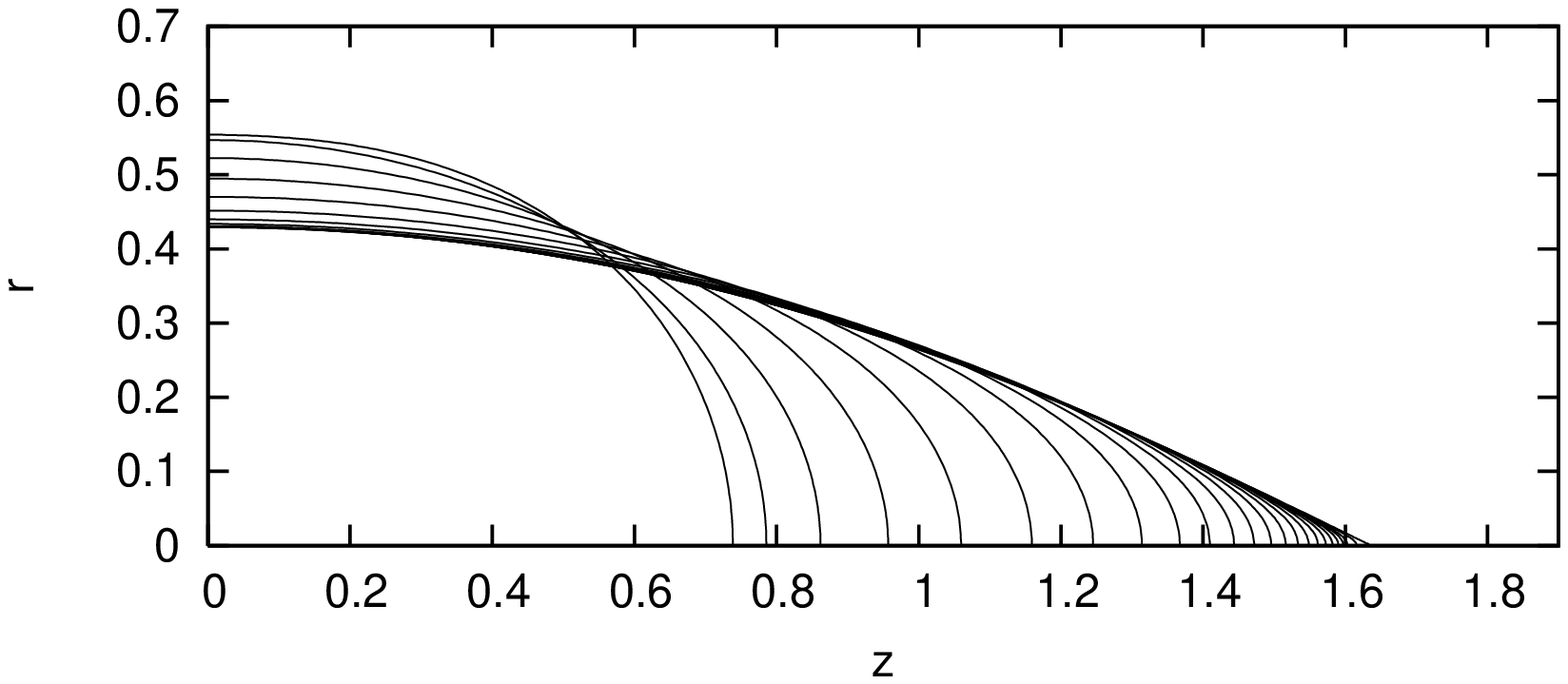}
\caption{Evolution of droplets with critical charge, with ratios of
viscosities $\protect\mu_1/\protect\mu_2=100$ (top) and $1$ (bottom). Notice
the tendency to develop conical tips.}
\label{fig1}
\end{figure}
The singularity occurs provided that the ratio of viscosities $\mu_1/\mu_2$
is large enough. Otherwise, the tips do not appear, and the drop tends to
separate into smaller droplets. Our numerical computations show that the
transition occurs between $\mu_1/\mu_2=0.1$ and $0.2$.

In figure 2 we show a close up in the region around the singularity. We
translate the $(r,z)$ coordinates laterally by an
amount $z_f=z(0,t)$ and then multiply them by the curvature at the
tip of the drop $k_f$. The figure shows that the local shape near the tip
is nearly invariant, and then by definition, the solution is self-similar.
\begin{figure}[tbp]
\centering
\includegraphics[width=0.75\textwidth]{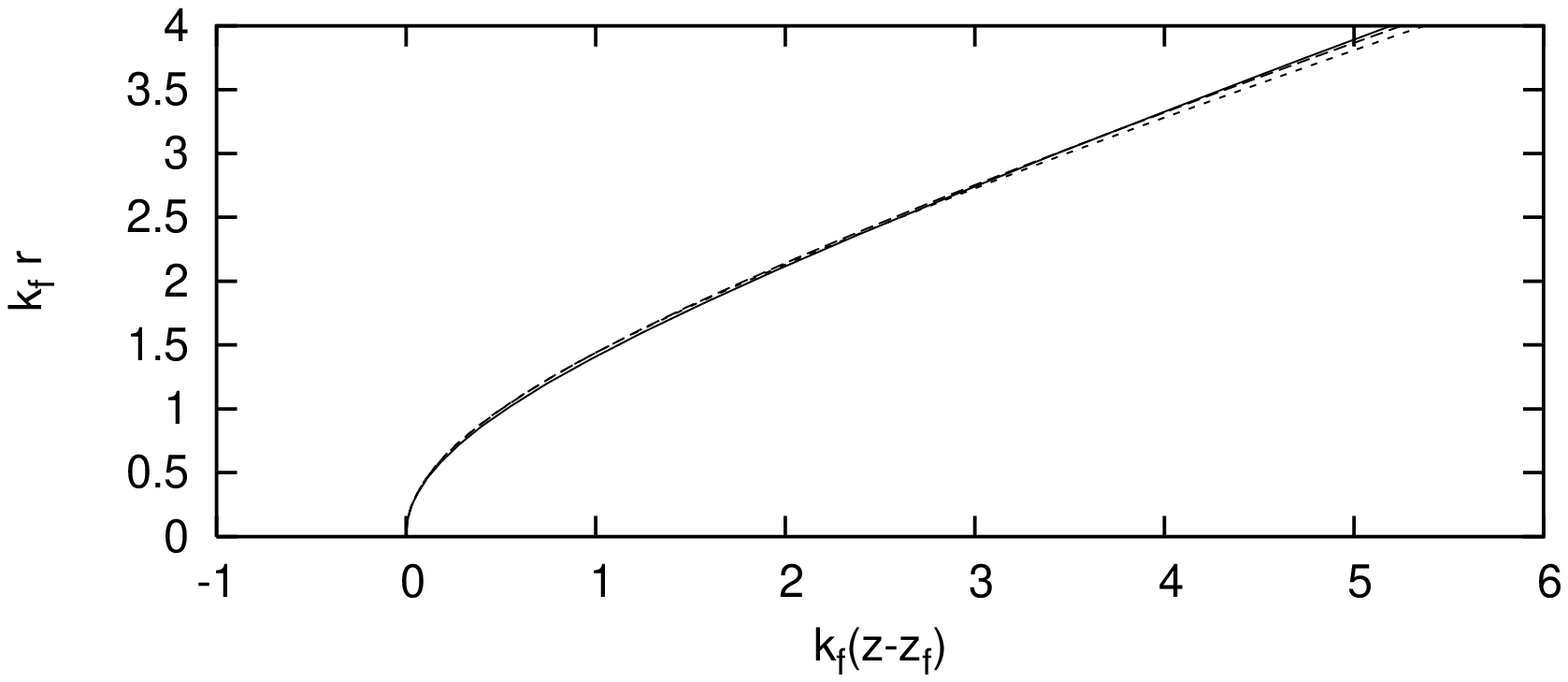}
\includegraphics[width=0.75\textwidth]{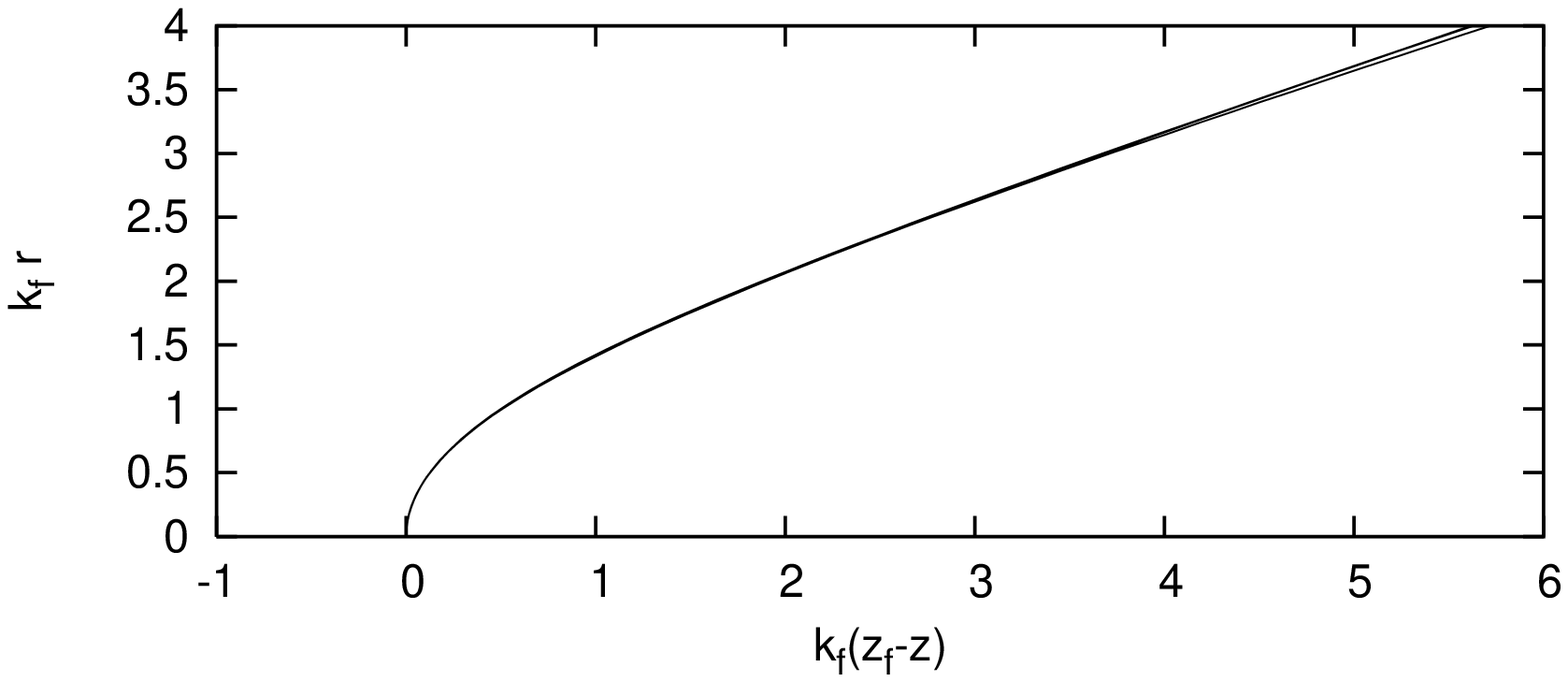}
\caption{Evolution of droplets with critical charge, with ratios of
viscosities $\protect\mu_1/\protect\mu_2=100$ (top) and $1$ (bottom). We
rescaled the free surfaces to make evident the self-similarity of the
shapes. The curvatures at the tip are $146.7, 92.1, 77.5 $ and $45.3$ for
the figure on the top, and $1739, 698, 112$ and $65$ for the figure at the
bottom. }
\label{fig2}
\end{figure}

In figure 3 we show plots of the velocity and charge density at the tip
(the point of maximum curvature), as a function of time.
This behaviour suggests a power-law or self-similar regime near the
time of the singularity. We also show the potential $V$ at the surface to
show that it does not diverge near the time of the formation of the tip.
The Reynolds number, defined as $v_f/k_f$, remains bounded, which indicates
that the hypothesis of the Stokes approximation are not violated near the
singularity time $t_0$.
\begin{figure}[tbp]
\centering \includegraphics[width=0.75\textwidth]{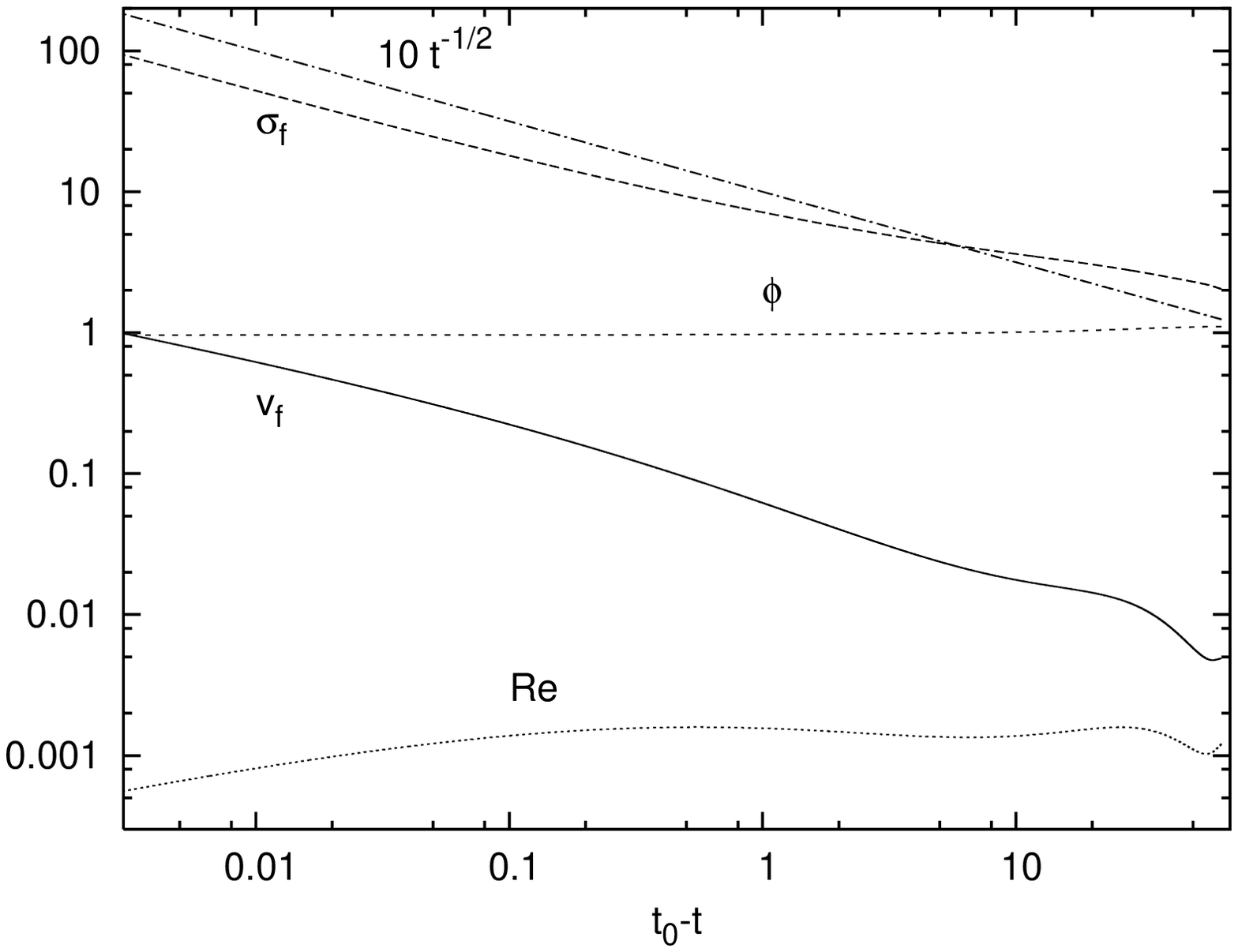}
\caption{Maximum charge density and fluid velocity at the surface of
droplets with critical charge, with ratios of viscosities
$\protect\mu_1/\protect\mu_2=1$.
The line on the top is a power law with exponent $-1/2$. The potential $V$
and the Reynolds number $Re$ are bounded during the evolution.}
\label{fig3}
\end{figure}

The angle of the conical tips depends weakly on the ratios of viscosities,
as shown in figure 4, and as mentioned above, it is much smaller than the
Taylor cone angle ($49.3^{o}$).
\begin{figure}[tbp]
\centering \includegraphics[width=0.75\textwidth]{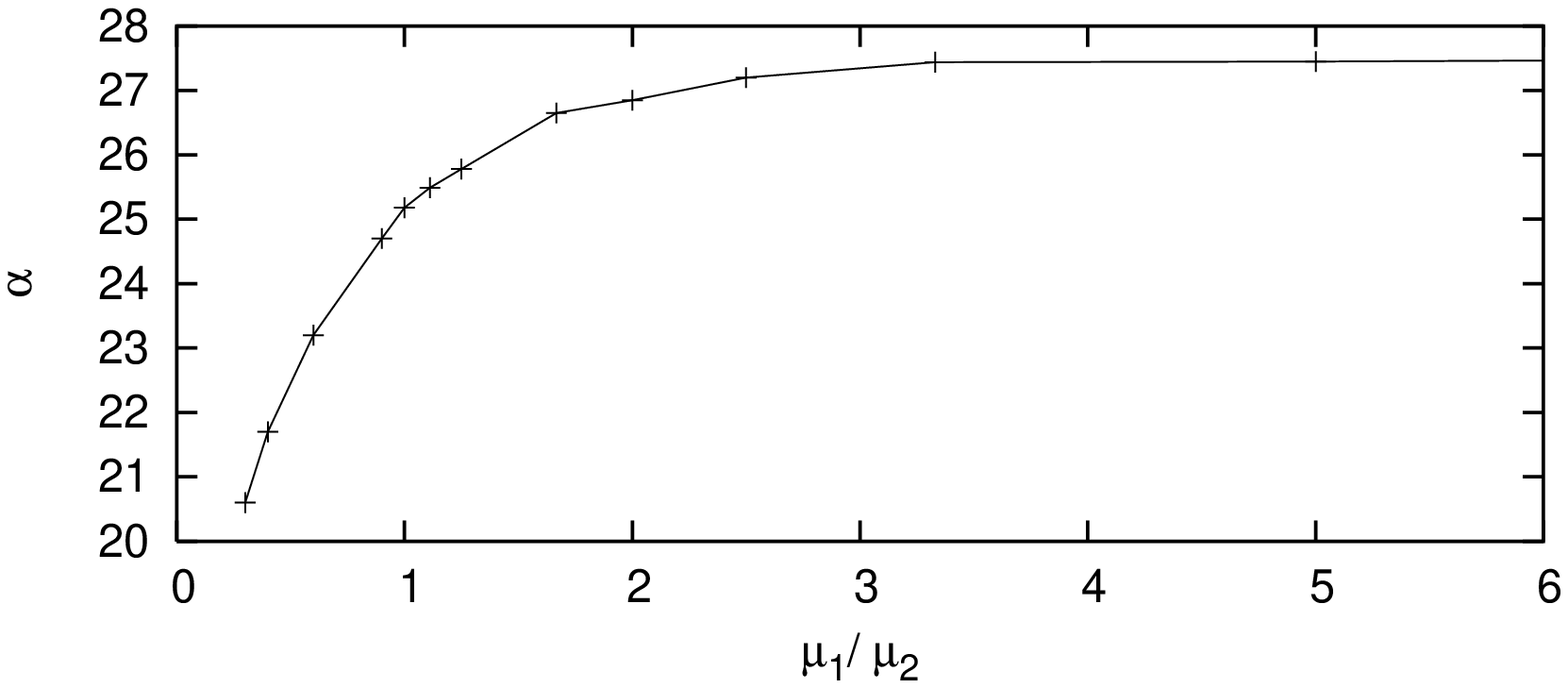}
\caption{Semiangle of the conical tips, as a function of the ratio of
viscosities $\protect\mu _{1}/\protect\mu _{2}$. }
\label{fig4}
\end{figure}

The numerical calculation does not allow us to explore much further
into the asymptotic regime $t\rightarrow t_0$. However we can argue with
a scaling argument that the similarity exponent
is $1/2$. We can also show that near the singularity, capillarity
forces are negligible while electric forces are dominant.

Let us place the origin of coordinates at a singularity point, where the
surface will eventually develop the vertex, and describe the free surface
with the following selfsimilar shape
\begin{equation}
z=(t_{0}-t)^{\alpha }f\left( \rho \right) \ ,\hspace{1cm}\rho
=r(t_{0}-t)^{-\alpha }  \label{zeta}
\end{equation}
where $t_{0}$ is the time of formation of the singularity and $\alpha >0$ is
a similarity exponent, that will be deduced from scaling arguments.

The numerical evidence indicates (figure 3) that the electric potential $V$
at the surface is a continuous
bounded function on the variable $t$, and $V$ has a non-zero limit as
$t\rightarrow t_{0}$.Then if we
assume a similarity solution for $V$ at the surface, it must have the form
\begin{equation}
V=\Phi \left( r(t_{0}-t)^{-\alpha }\right)
\end{equation}
and then
\begin{equation}
\sigma =-\varepsilon _{0}\frac{\partial V}{\partial n}=(t_{0}-t)^{-\alpha
}\Sigma \left( r(t_{0}-t)^{-\alpha }\right) ,
\end{equation}
and in particular, the charge density at the tip of the drop is proportional
to $1/(t_{0}-t)^{\alpha }$.

Since the mean curvature is given by
\begin{equation}
\kappa =\frac{1}{2}\left( \frac{1}{R_1}+\frac{1}{R_2}\right) \sim
K(t_0-t)^{-\alpha }\;\mbox{ as }\;t\rightarrow t_0,
\end{equation}
where the $\sim $ symbol means ``asymptotically proportional to'', and
\begin{equation}
\frac{\varepsilon_0}{2}\left( \frac{\partial V}{\partial n}\right)^2=
\frac{\sigma^2}{2\varepsilon_0}\sim C^{\prime }(t_0-t)^{-2\alpha
}\gg C(t_0-t)^{-\alpha },\mbox{ as }t\rightarrow t_0
\end{equation}
we can conclude that capillary forces are negligible in comparison with the
electrostatic forces near the singularity.
\begin{figure}[ht]
\centering \includegraphics[width=0.75\textwidth]{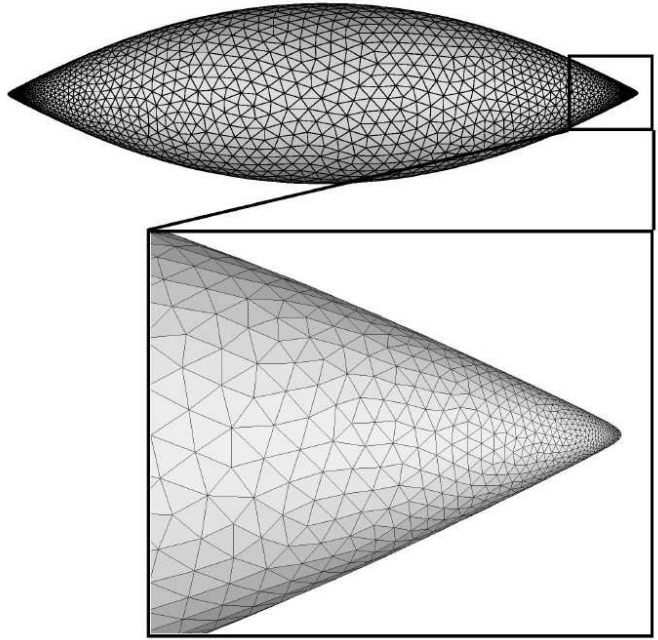} 
\caption{Evolution of a three dimensional drop: the singularity
also develops in a 3D flow and the surface develops the same conical tips.}
\label{fig5}
\end{figure}
\begin{figure}[ht]
\centering \includegraphics[width=0.75\textwidth]{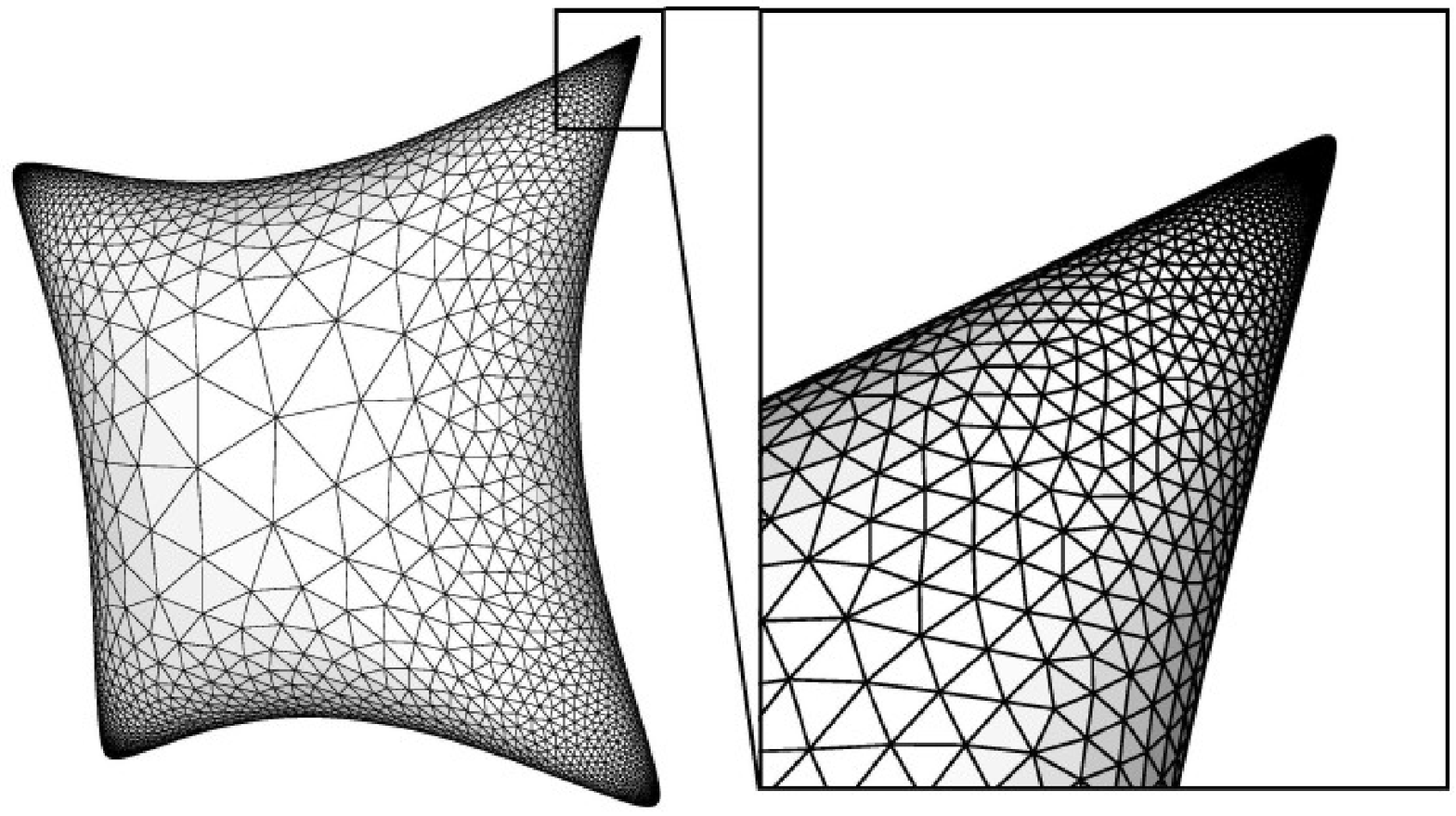}
\caption{Evolution of a non-symmetric drop, and the formation of
conical tips.} \label{fig6}
\end{figure}
Then, the balance of stresses at the surface yields
$\left\vert \mathbf{n}(T^{(2)}-T^{(1)})\mathbf{n}\right\vert \sim C^{\prime
}(t_0-t)^{-2\alpha }$,
from which we deduce typical pressure and velocity gradients
$p,u_{i,j}\sim O\left( (t_0-t)^{-2\alpha }\right)$,
and hence velocities
\begin{equation}
u_{i}\sim O\left( (t_{0}-t)^{-\alpha }\right) .  \label{orderu}
\end{equation}
By Eq. (\ref{zeta}), $\mathbf{r}_t\sim O\left( (t_0-t)^{\alpha
-1}\right) $, which together with the kinematic condition at the surface
$\mathbf{r}_{t}=\mathbf{u}$
and (\ref{orderu}) imply $\alpha -1=-\alpha $ and therefore
$\alpha =\frac{1}{2}.$ Hence the solutions of the Stokes system near the singularity
are expected to have the form
$p=(t_{0}-t)^{-1}P\left( \xi ,\rho \right)$ and $\mathbf{u}=(t_{0}-t)^{-\frac{1%
}{2}}\mathbf{U}\left( \xi ,\rho \right)$,
with $\rho =r(t_0-t)^{-\frac{1}{2}}$ and $\xi =z(t_{0}-t)^{-\frac{1}{2}}$.

By relaxing the assumption of circular symmetry, conical tips still are
developed. In order to show this, we implemented another numerical simulation
based on the boundary elements method with adaptive triangulated surfaces to
handle general three dimensional situations. Then we introduce a sphere
perturbed with the mode $Y_{2}^{0}$ as an initial condition with critical
charge, but, because we are approximating the shape of the surface with
triangles, the numerical approximation is not axially symmetric. The result
is in figure 5: the global shape is still axially symmetrical at the time
of singularity formation. This is strong evidence of the stability
of the self-similar solutions.
On the other hand, the formation of singularities does not appear to be restricted to
circularly symmetric flows: we found that they can also appear on asymmetric
configurations. In figure 6 we show the evolution of an initial surface
composed by an oblate ellipsoid with an asymmetric perturbation, with a
supercritical charge $Q=1.2Q_c$. It can be clearly seen the tendency to
develop a conical tip after the evolution. Moreover, the local shape of the tip is
circularly symmetric, and the angle of the tip is approximately $25^{o}$, of
the same order of magnitude as the angle on the symmetric shapes. This is a
further indication that the singularity angle is fairly insensitive to the
initial condition, and that it is smaller than Taylor's value.


\begin{thebibliography}{7}
\bibitem{Taylor} G. I. Taylor, Disintegration of water drops in an electric
field, \textit{Proc. Roy. Soc. London A} 280 (1964), 383-397.

\bibitem{Rayleigh} Lord Rayleigh, On the equilibrium of liquid conducting
masses charged with electricity, \textit{Phil. Mag. }\textbf{14} (1882),
184-186.

\bibitem{Duft} D. Duft, T. Achtzehn, R. M\"{u}ller, B. A. Huber and T.
Leisner, Rayleigh jets from levitated microdroplets, \textit{Nature}, vol.
\textbf{421}, 9 January 2003, pg. 128.

\bibitem{barrero} I. G. Loscertales, A. Barrero, I. Guerrero, R. Cortijo, M.
Marquez, A. M. Ga\~{n}an-Calvo, Micro/Nano Encapsutation Via Electrified
Coaxial Liquid Jets, \textit{Science}, Vol. 295, 5560 (2002), 1695-1698.

\bibitem{Pozrikidis} C. Pozrikidis, Boundary integral methods for linearized
viscous flow, Cambridge texts in Applied Mathematics, Cambridge University
Press, 1992.

\bibitem{Acrivos} J. M. Rallison, A. Acrivos, A numerical study of the
deformation and burst of a viscous drop in an external flow, \textit{J.
Fluid Mech.} \textbf{89} (1978), 191-200.

\bibitem{GCH} M. Gamero-Casta\~{n}o, V. Hruby, Electrospray as a source of
nanoparticles for efficient colloid thrusters, \textit{J. Prop. Power}
\textbf{17} (2001), 977-987.
\end{thebibliography}
\end{document}